# Antiferromagnetic magnon spintronic based on non-reciprocal and non-degenerated ultra-fast spin-waves in the canted antiferromagnet α-Fe$_2$O$_3$

Aya El Kanj[1], Olena Gomonay[2], Isabella Boventer[1], Paolo Bortolotti[1], Vincent Cros[1], Abdelmadjid Anane[1], Romain Lebrun[1,*]

[1] Unité Mixte de Physique, CNRS, Thales, Université Paris-Saclay, 91767 Palaiseau, France
[2] Institute of Physics, Johannes Gutenberg-University Mainz, 55128 Mainz, Germany
* Corresponding author: romain.lebrun@cnrs-thales.fr

**Abstract**
Spin-waves in antiferromagnets hold the prospects for the development of faster, less power-hungry electronics, as well as new physics based on spin-superfluids and coherent magnon-condensates. For both these perspectives, addressing electrically coherent antiferromagnetic spin-waves is of importance, a prerequisite that has so far been elusive, because unlike ferromagnets, antiferromagnets couple weakly to radiofrequency fields. Here, we demonstrate the detection of ultra-fast non-reciprocal spin-waves in the dipolar-exchange regime of a canted antiferromagnet using both inductive and spintronic transducers. Using time-of-flight spin-wave spectroscopy on hematite (α-Fe$_2$O$_3$), we find that the magnon wave packets can propagate as fast as 20 km/s for reciprocal bulk spin-wave modes and up to 6 km/s for surface-spin waves propagating parallel to the antiferromagnetic Néel vector. We finally achieve efficient electrical detection of non-reciprocal spin-wave transport using non-local inverse spin-Hall effects. The electrical detection of coherent non-reciprocal antiferromagnetic spin waves paves the way for the development of antiferromagnetic and altermagnet-based magnonic devices.

**MAIN TEXT**

**Introduction**
Spin wave dynamics in antiferromagnets hold the prospect of magnonic devices operating at the sub-THz frequencies(*1–3*) with a large group velocity (> 10 km/s ) by benefiting from their strong exchange field and quadratic spin-wave dispersion [4,5]. In this context, antiferromagnetic spin-waves in the long and short (including dipole-exchange modes) wave-length limits have been extensively investigated theoretically already some decades ago(*6–9*). For magnonic devices, one of the most basic actions to be realized is to be able to electrically excite and detect the corresponding fast spin waves. However, up to now, there are no experimental observations of propagating properties of spin waves in AFMs, in both direct and reciprocal space. Indeed, contrary to their counterparts ferromagnets, in which large stray fields allow to detect inductively the spin wave dynamics relatively straightforwardly, such dipolar fields in antiferromagnets are zero or largely negligible. Beyond their key role in spin wave detection, the non-compensated dipolar fields also provide some of the unique features such as non-reciprocity, magneto-static spin-waves(*10, 11*), Bose-Einstein condensation(*12, 13*)) of conventional ferromagnet-based magnonic devices. Due to the bulk Dzyaloshinskii-Moriya interaction(*14, 15*), canted antiferromagnets are anticipated to present more pronounced dipole-exchange spin-wave modes in the small wave vector **k** region (< 6 rad/μm)(*16–19*) as required to facilitate their observations using standard inductive detection. Recently, many of these canted antiferromagnet materials, such as hematite and orthoferrites, have also been identified as altermagnets(*20, 21*), a new class of magnetic materials with opposite spin-sub-lattices, and a nearly vanishing compensated magnetic order but at the same time a broken T-symmetry leading to spin-splitting in the momentum space. Such a lifted degeneracy of the electronic spin and magnon band structures shall enable to open to antiferromagnets, the same rich physics of spin current transport and spin wave dynamics(*22*) than in ferromagnets(*23*). In this sense, insulating canted antiferromagnets such as hematite, the material to be studied here, or orthoferrites, with resonance frequencies ranging from 10 to 600 GHz(*24–26*), DMI fields from 1 to 20 T(*27*), and low magnetic damping(*25, 28, 29*), are thus prime candidates to develop the field of antiferromagnetic and alter-magnonics. In the last decade, research in spintronic proposed various approaches to enable the detection and the manipulation of antiferromagnetic spin-waves using spin-to-charge phenomena(*30–32*). Until now, electrical detection based on inverse spin-Hall effect was achieved only for the uniform mode (**k** = 0) with generated voltage amplitudes as low as tens of nV in both colinear(*2, 3*) and canted antiferromagnets(*33, 34*).

In this article, we successfully identify magneto-static spin-waves for low **k** vector (0.1 – 2.3 rad/µm) in hematite (α-$Fe_2O_3$). To this aim, we first used spin-wave spectroscopy between two inductive transducer antennas allowing us to detect these AFM spin waves after propagating on a distance of more than 10 µm. Using time of flight spin-wave spectroscopy(*35*), we evidence the presence of different spin-wave packets with very large group velocities ranging from 5 to 30 km/s. In addition, we report a strongly lifted degeneracy of the bulk spin-wave band for **k** perpendicular (⊥) or parallel (//) to the antiferromagnetic order **n**, with a separation larger than 1 GHz at k = 0.6 rad/µm and demonstrate clearly the non-reciprocal character of spin-wave modes for **k** // **n**, a highly interesting feature for the development of antiferromagnetic magnonics. Lastly, we achieve electrical detection of the non-reciprocal antiferromagnetic spin-waves with a platinum based metallic transducer (through the inverse spin-Hall effect) with µV output voltage, as in ferromagnets like YIG(*36, 37*). Our observations evidences that spintronic transducers represent a promising alternative to detect antiferromagnetic spin-waves with reduced dipolar fields.

**Results**
**Lifting of magnon degeneracy in canted antiferromagnets**
We excite and detect propagating spin waves in c-plane oriented single crystals of the canted antiferromagnet α-$Fe_2O_3$ (*14, 38, 39*) by means of propagative spin wave spectroscopy(*40*) (cf. **Fig. 1 (a)**). First, we measure with a Vector Network Analyzer (VNA) the reflected $L_{11}$ and transmission $L_{21}$ parameters between two inductive antennas that predominantly excite spin-wave with **k** vector of 0.6 rad/µm (see more details in Methods and **Suppl. Mat. 1**(*41*)). We succeed in detecting spin-wave propagation for an edge to edge distances as large as 14 µm. Such a long-distance coherent transport of antiferromagnetic spin-waves is in line with the recently reported micrometer-long magnon spin-diffusion length and the ultra-low magnetic damping of hematite(*28*). Due to the small canted moment **m** of α-$Fe_2O_3$ ($M_s$ ≈ 3 emu/cm³(*38*)), the direction of the antiferromagnetic order **n** can be oriented perpendicular to the applied field **H** with fields as low as 50 mT (*38*). Indeed, hematite is one of the rare easy-plane antiferromagnet at room temperature and we specifically chose the sample to have the magnetic easy-plane parallel to the surface. This property allows us to investigate spin-wave propagation for an antiferromagnetic order **n**, either parallel or perpendicular to the spin-wave vector **k**. As described here after, these measurements lead to the observation of strikingly different behaviors for **k** // **n** and **k** ⊥ **n**. First, for **k** ⊥ **n** (i.e., **H** // **k**), we observe as shown in **Fig. 1 (b)** two close spin-wave branches, around 19 GHz at 150 mT. This could be associated to slightly non uniform anisotropies of the sample(*28, 42*). For **k** // **n** (i.e., **H** ⊥ **k**), we observe a main spin-wave mode (blue line) as shown in **Fig. 1 (c)** along with several spin-wave branches at slightly higher frequencies (represented by orange and green lines). These features could indicate the presence of magneto-static modes(*16, 19, 28, 43*), and we will discuss later how to identify them. It is to be noticed that the lowest spin-wave branch (blue branch) for **k** // **n** follows a similar frequency dispersion as for **k ⊥ n**, but is always higher in frequency by about 1 GHz. We also see that the signal amplitude strongly varies below 50 mT, this is due to, as mentioned above, the reorientation of the Néel vector **n** and canted moment **m**(28, 44) in this low field range.

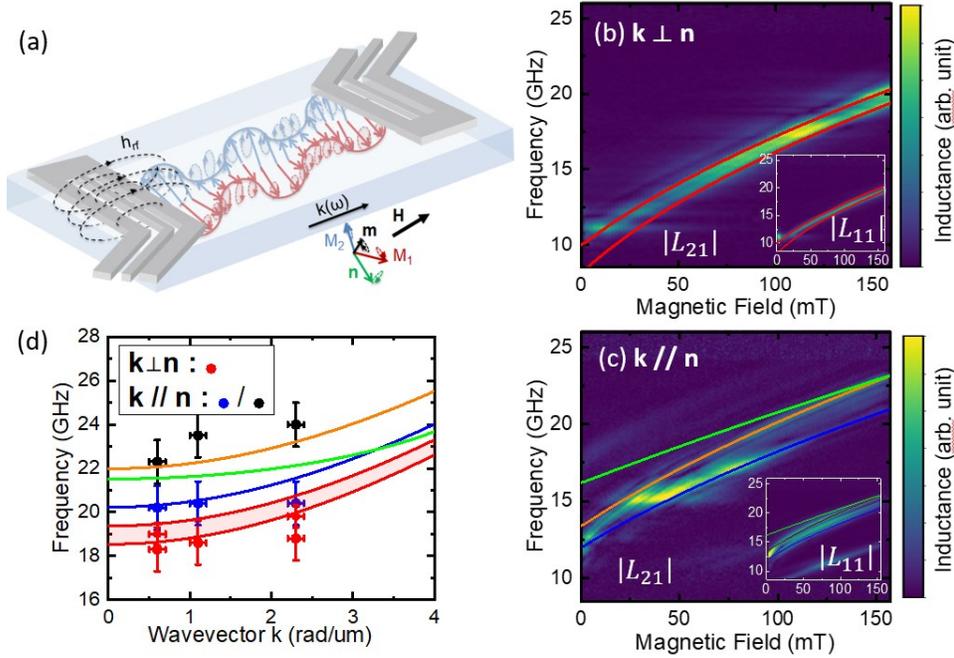

**Figure 1. Spin-wave transport in the canted antiferromagnet α-Fe₂O₃.** (a) Schematics of the setup. The net sub-lattice magnetization M₁ and M₂ have strongly elliptical trajectory, oscillating mainly in the sample plane (easy-plane) with only a small opening angle in the out-of-plane direction. **n** and **m** respectively correspond to the Néel vector and the canted moment dynamics. **n** is linearly polarized in the sample plane whilst **m** is elliptically polarized. The static net moment **m** is aligned along the applied field **H**, and the Néel vector **n** is perpendicular to it. (b-c) Spin wave transmission measurement showing the transmitted amplitude |L₂₁| for (b) **k ⊥ n** and **k // n** for (c) at k ≈ 0.6 rad/μm. (Red and blue lines correspond to fits using the theoretical bulk spin-wave equations for **k // n** and **k ⊥ n**. Orange and green lines respectively correspond to a modelling of the high-frequency spin-wave branch for **k // n** assuming a bulk or a surface mode (see **Suppl. Mat. 7** and Refs. (16, 17, 19)). Insets show the amplitude of the reflected signal |L₁₁|. (d) Magnon branch dispersion for **k // n** and **k ⊥ n** at a magnetic field of 140 mT. (Blue and red lines correspond to the theoretical bulk spin-wave branches **k // n** and **k ⊥ n** respectively) using the fitting in magnetic field).

To understand the origin of this anisotropic magnon transport, we measure in **Fig. 1 (d)** the spin-wave dispersion for **k // n** (black and blue points) and **k ⊥ n** (red points) by performing spin-wave spectroscopy at different **k** vectors using several transducer designs (see **Suppl. Mat 1**(41)). We clearly observe the persistence of well-separated magnon branches for the two configurations, with always higher frequencies for **k // n** (see **Suppl. Mat 3**(41)). Indeed, such a lifted degeneracy of the magnon dispersion is not expected from the standard degenerated linear dispersion reported for antiferromagnets. To go beyond, other regimes should be considered, such as the dipole-exchange regime of canted antiferromagnets, which in our knowledge has not been yet explored experimentally. As for the difference in spin-wave frequency between the two configurations **k ⊥ n** and **k // n**, some theoretical models(16, 17) do predict that the bulk spin-wave dispersion should vary between these two cases. The refined expression of the bulk spin-wave bands (see **Suppl. Mat 5**(41)) in the dipole-exchange regime leads to a frequency difference $\Delta f_{SW} = f_{\mathbf{k} // \mathbf{n}} - f_{\mathbf{k} \perp \mathbf{n}} =$
$\sqrt{f_{10}^2 + \frac{4\pi M_s}{H_{ex}}\left(\frac{\gamma}{2\pi}\right)^2 (H + H_{DMI})^2} - f_{10}\left(1 + \frac{4\pi M_s}{H_{ex}}\right)$ with $f_{10}$ the frequency gap for the lowest magnon mode(33, 42), $\gamma$ is gyromagnetic ratio, $H_{ex}$ is exchange field and $H_{DMI}$ is the Dzyaloshinskii-Moriya field. Using the material parameters of hematite(41), we estimate $\Delta f_{SW} \approx$ 0.5 – 1 GHz for small **k** vectors (10 rad/μm) which agrees with the observation that the spin-wave frequencies are higher for **k // n** than for **k ⊥ n**. We can fit the frequency of the bulk spin-wave modes versus fields for these two configurations (see respectively red and blue in **Fig. 1 (b-c)**). This result evidences the importance of magneto-static interactions in the spin-wave dynamics of canted antiferromagnets at small **k** vectors (< 10 rad/µm). However, these models cannot explain the presence of the higher frequency spin-wave branches present for **k** // **n**.

**Time of flight of surface and bulk antiferromagnetic spin-waves**
In order to get more insights about the properties of these propagating AFM spin-waves, we analyze their amplitude and their phase in more details for **k** // **n** and **k** ⊥ **n**. We restrict our analysis for magnetic fields above 50 mT to ensure that the Néel vector **n** is always strictly perpendicular to the **H**. In **Fig. 2 (a)**, we present the imaginary part of the transmitted spin-wave spectra $L_{21}$ for **k** ⊥ **n**. As shown in **Fig. 2(b),** we observe the expected oscillatory behavior of the phase delay $\varphi = kD_{ant}$ (with $D_{ant}$ the distance between the two transducer antennae) accumulated by the spin wave during its propagation. From these oscillations, the spin-wave group velocity $v_g = \frac{\partial f}{\partial k} \sim D_{ant} \Delta f$ can be extracted from the periodicity of phase oscillations $\Delta f$. However as shown in the black curve of **Fig. 2 (b)**, the envelop of the signal shows the presence of more than one spin-wave packet. Those are due to both wide k-bandwidth of our antenna ($\partial k \approx 0.2 - 1$ rad/µm, see **Suppl. Mat. 1**(*41*)), and potentially to propagating spin-waves with non-uniform thickness profile in our 500 µm thick film. To access the group velocity of each spin-wave packets, we perform time gating VNA measurements(*35*) with different time intervals (see **Suppl. Mat 2**(*41*)). As shown in **Fig. 2(b)**, we detect the main (and fastest) spin-wave packets in less than 1 ns of travelling time for an edge-to-edge distance of 14 µm between the antennas, indicating a minimum group velocity > 14 km/s. Note that a travelling time of 1 ns (no spin-wave packet after 2 ns) is compatible with the group velocity that can be extracted from the phase oscillations $\Delta f$ in **Fig. 2 (c)**, that lies in an average value of around 20 km/s over the measured field range. We emphasize that group velocities as our reported value, represent a record velocity for spin-waves in a magnonic device.

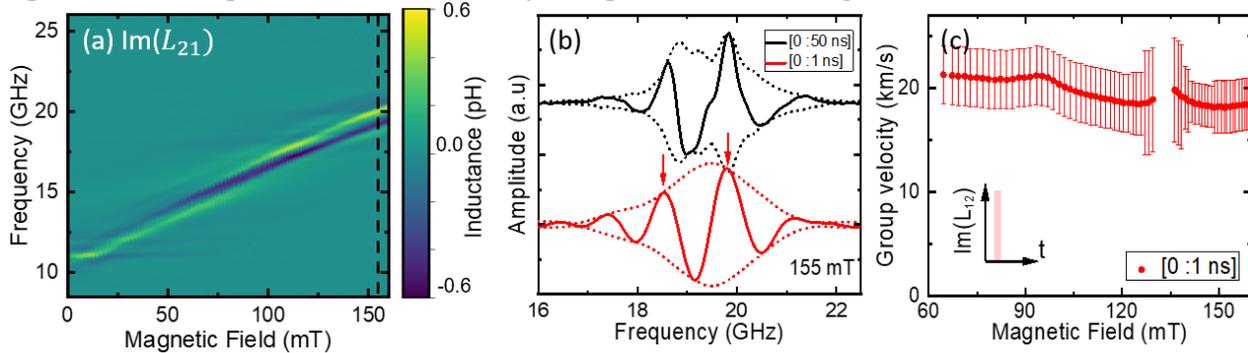

**Figure 2. Ultra-fast antiferromagnetic spin-waves for k ⊥ n revealed by time gated VNA measurements.** (a) Imaginary part of the transmitted spin-wave Im($L_{21}$) as a function of field without time gating. (b) Example of Im($L_{21}$) spectra of the full spin-wave signals (time gate of [0:50 ns], black) and of the main spin-wave packet (time gate of [0:1 ns], red) for H = 155 mT (No spin-wave signal is detected after 10 ns). For an edge-to-edge antenna distance of 14 µm, oscillations (red) indicate a spin-wave group velocity > 14 km/s. Dotted lines correspond to the signal envelops. (c) Group velocity of the main spin-wave packet for a time gating of [0:1 ns]. Error bars are defined as the noise level from the imaginary part of the transmitted inductance $L_{21}$.

Contrary to ferromagnets in which the group velocity at small **k** scales with the magnetization saturation $M_s$(*45*), the group velocity $v_g$ in both collinear and canted antiferromagnets is proportional to $H_{ex}$(*46*). Thus, in antiferromagnets, it results that the group velocity can reach tens of km/s as observed in the present work(*4*), or that domain-wall velocity can be a few km/s as in orthoferrites(*5*). Note that the observed large spin-wave velocity is also in agreement with the value estimated from the experimental slope of the spin-wave dispersion $\frac{\partial f}{\partial k}$ presented in **Fig. 1 (d),** which also correspond to spin-wave velocity larger than 10 km/s.

In **Fig. 3**, we present the spin-wave propagating properties in the geometry **k** // **n**. As mentioned before, spin-wave branches separated by a few GHz can be observed in this case. Whilst the first one can be associated to bulk-spin-wave, the higher frequency ones could correspond to the predicted surface spin-wave modes or hybrid surface-bulk modes(*19, 42, 43, 47*) (see **Suppl. Mat**

5(*41*)). In **Fig. 3,** we thus present time gating measurements to independently access these different spin-waves modes. As for the configuration **k ⊥ n**, we observe that the first (and fastest) spin-wave packet travels in less than 1 ns and exhibit a group velocity of about 20 km/s (see **Fig. 3 (c)**). It slightly increases with field, leading to larger phase oscillations $\Delta f$ that become difficult to extract above 120 mT. Surprisingly, the higher frequency spin-wave modes (see blue curves in **Fig. 3(b-c))** propagates more slowly but still travel in less than 10 ns. As they are close in frequencies, and have similar travelling time, we only determine average group velocities to be around 6 km/s.

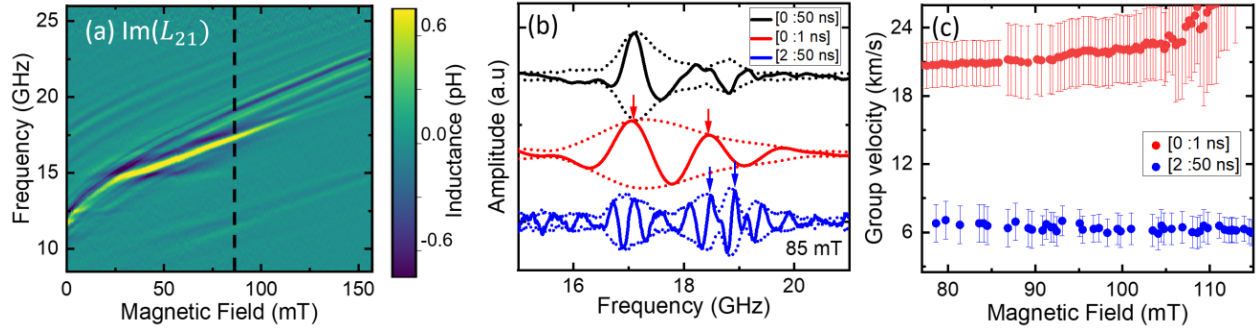

**Figure 3. Ultra-fast antiferromagnetic spin-waves for k // n.** (a) Imaginary part of the transmitted spin-wave Im($L_{21}$). (b) Exemplary spectra of Im($L_{21}$) for H = 85 mT for different time gating, [0:50 ns]: full spin-wave signals (black), [0:1 ns]: first spin-wave packet (red), [2:50 ns]: secondary spin-wave packets (blue). Dotted lines correspond to the signal envelops. (c) Group velocity of the different spin-wave packets. The antenna show a **k** selectivity centered around 0.6 rad/μm (see design in Methods). Error bars are defined as the noise level from the imaginary part of the transmitted inductance $L_{21}$. Higher frequency modes propagate around 3 times slower with around 8 km/s than the bulk mode, with around 20 km/s.

### Non-reciprocal spin-wave transport

Non-reciprocity is a key property for many spin-wave analog devices (such as circulators). It has been widely studied in ferromagnets in presence of surface spin-wave modes(*48*) but up to our knowledge only predicted in antiferromagnets(*7, 8*). Here, we thus investigate the potential non-reciprocity of the high frequency spin-wave packets for **k // n**. In **Fig. 4 (a-b)**, we present the amplitude of the transmitted spin-wave packets $|L_{21}|$ for positive and negative fields respectively. We use a time gating of [2:50 ns] to select the high-frequency spin-wave modes (see **Fig. 3 (a)**). As seen in **Fig. 4 (c)**, we do not observe a sizeable frequency shift between positive and negative magnetic fields. However, as far as the spin-wave amplitude is concerned, we find a clear non-reciprocity for two out of the three spin-wave modes. As shown in **Fig. 4 (d),** we observe for negative magnetic fields a reduction by about a factor 2 of the red mode and even the absence of the blue mode. This non-reciprocal behavior is confirmed by measuring different amplitudes for $|L_{21}|$ and $|L_{12}|$ parameters (see **Suppl. Mat. 6**(*41*)). These results are signatures of surface spin wave modes propagating with opposite directions at the two surfaces of the sample for **k // n**, which are expected also in case of an antiferromagnet(*7, 8, 19*). On the contrary, for **k ⊥ n,** we measure similar spin-wave amplitudes for the different spin-wave packets between positive and negative fields, and between $L_{12}$ and $L_{21}$ parameters (see **Suppl. Mat. 5**(*41*)), indicating a reciprocal behavior in this configuration(*7, 8*). To understand in more details the symmetry of these spin-waves and how they decay within the AFM requires further theoretical investigation and is beyond the scope of this work. This can be done by using either the canted antiferromagnet approach (see **Suppl. Mat. 5**(*41*)) or the altermagnet formalism.

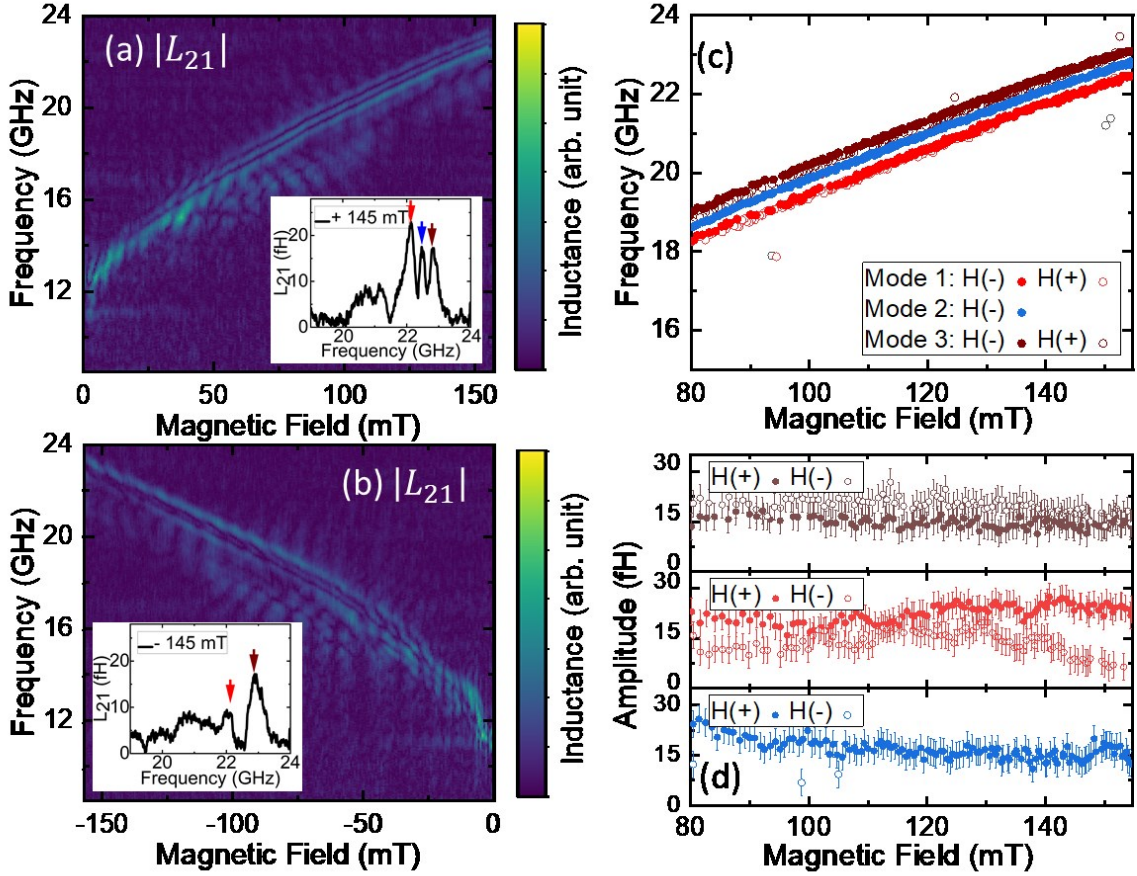

**Figure 4. Non-reciprocal spin-wave for k // n.** (a-b) Absolute value of the transmitted spin-wave spectra $|L_{21}|$ with a time gating of [2:50 ns], for positive fields in (a) and negative in (b). Insets shows exemplary spectra for H respectively of + 145 mT and – 145 mT. Arrows indicates the position of the 3 different modes. (c) Frequency and (d) amplitude of the three main spin-wave modes for negative and positive fields. Error bars are defined as the noise level from the transmitted inductance $L_{21}$.

### Inverse spin-Hall detection of non-reciprocal spin-wave propagation

A key challenge in magnonic devices is the amplitude of the output voltage generated by the propagating spin-waves and efficient alternatives to standard inductive transducers(*49*) are still lacking. This challenge is even amplified in antiferromagnetic materials, given the reduced generated stray field. Here, we finally achieve efficient electrical detection of the propagating spin-waves through the surface sensitive inverse spin-Hall effect using a platinum based metallic transducer (see sketch in **Fig. 5 (a)**). As seen in **Fig. 5 (b)**, we observe a sign reversal of the generated DC voltage for positive and negative fields, indicating its spin-pumping nature. Another important feature is the strong asymmetry (about 40%) of the output voltage, which indicates the non-reciprocity of the detected spin-waves in line with our previous observations and evidences their presence at the surface of the crystal. We also study the angular dependency of the inverse spin-Hall voltage in **Fig. 5 (c-d)**. In **Fig. 5 (d)**, we observe that the resonance field at 17 GHz is larger for **k // n** than for **k ⊥ n**, which confirms with the results from **Fig. 1**. Furthermore, we notice in **Fig. 5 (c)** that the detected output voltage follows an asymmetric $(A\cos\theta^2 + B)\sin\theta$ dependency, with maxima for external magnetic field applied at 45° and 135°, from the transducer direction. This feature is in accordance with an excitation efficiency of the inductive transducers, which follows a $(A\cos\theta^2 + B)$ law (see **Suppl. Mat. 6**), and inverse spin-Hall detection, which follows a standard $\sin\theta$ law(*33, 37*). The additional asymmetry arises from the spin-wave non-reciprocity discussed in the previous section. One should mention that the shape of the output voltage peak can change towards high power due to nonlinear effects coming into play (arising from the ultra-low damping of hematite) that would require further study. By comparing the voltage amplitude for two distances between the injector and the detector, we also extract an attenuation length of about 3-4 µm. As shown in the inset of **Fig. 5 (b)**, we measure output inverse spin-Hall voltages in the microvolt range, whilst the excitation frequency being one order of magnitude larger than in ferromagnets(*36*, *37*). This further evidences that spin-pumping effects represent a promising tool to detect the spin-wave dynamics in antiferromagnets, and favorize their integration in magnonic devices.

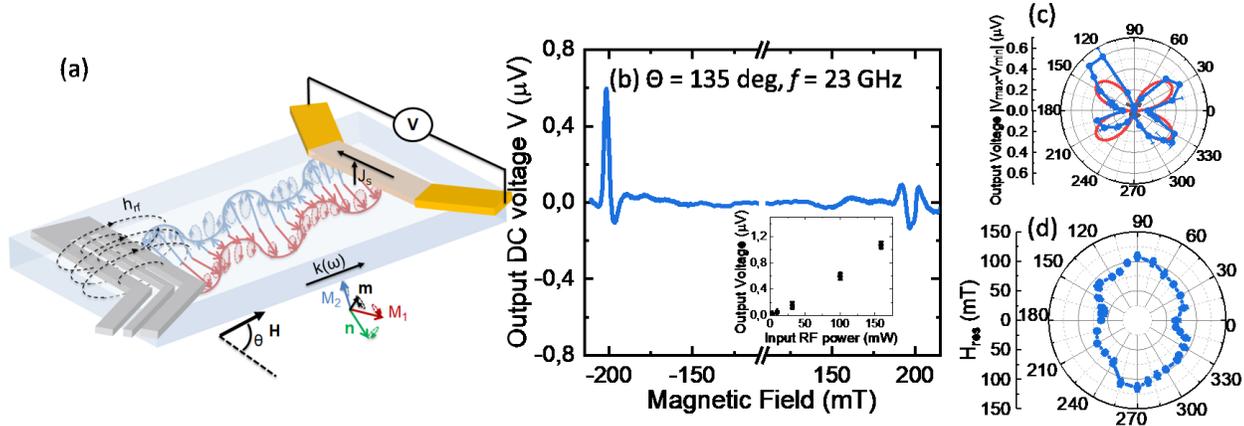

**Figure 5 Detection of antiferromagnetic spin-wave by inverse spin-Hall effects** (a) Sketch of the devices. An inductive transducer as in Fig. (1) is employed to excite the coherent antiferromagnetic spin-wave and a platinum based detector is used to detect the DC inverse spin-Hall voltage generated by the propagation spin-waves. (b) Example of inverse spin-Hall spectra for a magnetic field applied at θ = 135° from the inductive transducer (with an input power $P_{rf}$ = + 16 dBm). Inset: Power dependency of the peak to peak output voltage for $f$ = 17 GHz and θ = 45 °. (c) Angular dependency of the DC inverse spin-Hall voltage $V_{pp}$ for a 14 µm distance between the injector and the detector for $f$ = 17 GHz. Red line corresponds to the fit with a $(A \cos \theta^2 + B)|\sin \theta|$ (d) Angular dependency of the resonance field for $f$ = 17 GHz.

**Discussion**

We thus electrically detect by both inductive and spintronic transducers the presence of non-degenerated and non-reciprocal spin-waves in the dipolar-exchange regime of a canted antiferromagnet, with record group velocities (of about 20 km/s) and micrometers propagation distances. We can well model the presence of a bulk spin-wave frequency band of a few GHz with lifted degeneracy for **k** ⊥ **n** and **k** // **n**, which is anticipated to be a generic feature for canted antiferromagnets at low k vectors. Furthermore, for **k** // **n**, we observe the co-existence of non-reciprocal with reciprocal spin-wave modes in **Fig. 4**. This non-reciprocal behavior is even enhanced at larger **k** (see **Suppl. Mat 4**(*41*)). Without considering coupling between surface and bulk modes, we can theoretically determine the frequency of the antiferromagnetic surface spin-wave modes to be $f_{sur} = \frac{f_{10}^2 + \left(\frac{ck}{2\pi}\right)^2}{\frac{\gamma}{\pi}(H+H_{DMI})} + \frac{\gamma}{4\pi}\left(1 + \frac{4\pi M_S}{H_{ex}}\right)(H + H_{DMI})$. Using the material parameters of hematite, we would thus expect spin-wave surface modes at around 32 GHz at 100 mT for k ≈ 0.6 rad/µm. This value is definitively larger than our experimental observations shown in **Fig. 4**, and furthermore the required stability conditions to be localized on the surfaces are not fulfilled (See **Suppl. Mat. 5**(*41*)). Hence, the frequency of stable surface modes at around 20 GHz can only be fitted with an unrealistic phenomenological effective DMI field ($H_{DMI}$ ≈ 1.3 T, see green lines in **Fig. 1 (b)**). However, similarly to what happens in thick ferromagnets(*47*), bulk and surface spin-waves can also strongly hybridize in a single crystal, leading to spin-wave modes with mixed properties. This could explain the presence of frequency spin-wave modes close in frequency with either non-reciprocal or reciprocal spin-wave behaviors as observed here in **Fig. 4**. Overall, our findings can only partially be modelled with a standard theory of antiferromagnetic spin-waves (developed in more detailed in **Suppl. Mat 5**(*41*)). Thus, further theory works would require to investigate in more details the mode interaction and the hybridization depending on the system geometry(*16*, *17*, *19*), together with the altermagnetic character of α-$Fe_2O_3$(*20*, *21*, *23*). The large spin-pumping signals generated by the propagating antiferromagnetic spin-waves also provides a promising tool to access their dynamics in both single crystals and thin films. One should notice that the large inverse spin-Hall voltage can be linked with the amplitude of the spin-Hall magnetoresistance(*39*, *50*) reported in bilayers of Hematite/Platinum, as large as in bilayers of YIG/Platinum(*51*). Given the low magnetic damping of other few orthoferrites, the material class of canted antiferromagnets demonstrate all its potential for establishing a new research field around antiferromagnetic and alter-magnonics, with a lot of opportunities for high frequency magnonics.

**Note:** In the preparation of this manuscript, we became aware of two recent works on spin-wave spectroscopy in hematite in which the authors also observed large group velocities of tens of km/s and long propagation distances(*52*, *53*). Our work evidences that, due to the presence of Dzyaloshinskii-Moriya field, the spin-wave dispersion in this canted antiferromagnet or altermagnet is non-trivial compared to the standard description of an antiferromagnet, being strongly non-degenerated due to magneto-static interaction at small **k** vectors, and can show reciprocal and non-reciprocal behaviors. The propagating surface spin-wave can then be efficiently detected using spintronic transducers using inverse spin-Hall effects.

**References**
1. T. Kampfrath *et al.*, *Nat. Photonics*. **5**, 31–34 (2011).
2. J. Li *et al.*, *Nature*, 1–5 (2020).
3. P. Vaidya *et al.*, *Science*. **368**, 160–165 (2020).
4. J. R. Hortensius *et al.*, *Nat. Phys.* **17**, 1001–1006 (2021).
5. V. G. Bar'yakhtar, B. A. Ivanov, M. V. Chetkin, *Sov. Phys. Uspekhi*. **28**, 563–588 (1985).
6. R. E. Camley, *Phys. Rev. Lett.* **45**, 283–286 (1980).
7. B. Lüthi, D. L. Mills, R. E. Camley, *Phys. Rev. B*. **28**, 1475–1479 (1983).
8. B. Lüthi, R. Hock, *J. Magn. Magn. Mater.* **38**, 264–268 (1983).
9. R. L. Stamps, R. E. Camley, *J. Appl. Phys.* **56**, 3497–3502 (1984).
10. B. A. Kalinikos, A. N. Slavin, *J. Phys. C Solid State Phys.* **19**, 7013–7033 (1986).
11. J. R. Eshbach, R. W. Damon, *Phys. Rev.* **118**, 1208–1210 (1960).
12. S. O. Demokritov *et al.*, *Nature*. **443**, 430–433 (2006).
13. B. Divinskiy *et al.*, *Nat. Commun.* **12**, 6541 (2021).
14. T. Moriya, *Phys. Rev.* **120**, 91–98 (1960).
15. I. Dzyaloshinsky, *J. Phys. Chem. Solids*. **4**, 241–255 (1958).
16. W. Jantz, W. Wettling, *Appl. Phys.* **15**, 399–407 (1978).
17. V. I. Ozhogin, *JETP Lett.* **21**.
18. R. Orbach, *Phys. Rev.* **115**, 1189–1193 (1959).
19. V. V. Tarasenko, V. D. Kharitonov, *JETP Lett.* **33** (1971).
20. L. Šmejkal, J. Sinova, T. Jungwirth, *Phys. Rev. X*. **12**, 031042 (2022).
21. L. Šmejkal, J. Sinova, T. Jungwirth, *ArXiv210505820 Cond-Mat* (2021) (available at http://arxiv.org/abs/2105.05820).
22. L. Šmejkal *et al.*, Chiral magnons in altermagnetic RuO2 (2022), , doi:10.48550/arXiv.2211.13806.
23. L. Šmejkal, J. Sinova, T. Jungwirth, *Phys. Rev. X*. **12**, 040501 (2022).
24. M. Białek, A. Magrez, A. Murk, J.-Ph. Ansermet, *Phys. Rev. B*. **97**, 054410 (2018).
25. S. Das *et al.*, *Nat. Commun.* **13**, 6140 (2022).
26. G. F. Herrmann, *Phys. Rev.* **133**, A1334–A1344 (1964).
27. D. Treves, *Phys. Rev.* **125**, 1843–1853 (1962).
28. R. Lebrun *et al.*, *Nat. Commun.* **11**, 6332 (2020).
29. M. Białek, J. Zhang, H. Yu, J.-Ph. Ansermet, *Appl. Phys. Lett.* **121**, 032401 (2022).
30. R. Cheng, M. W. Daniels, J.-G. Zhu, D. Xiao, *Sci. Rep.* **6**, 24223 (2016).

31. I. Proskurin, R. L. Stamps, A. S. Ovchinnikov, J. Kishine, *Phys. Rev. Lett.* **119** (2017), doi:10.1103/PhysRevLett.119.177202.
32. R. Cheng, D. Xiao, A. Brataas, *Phys. Rev. Lett.* **116**, 207603 (2016).
33. I. Boventer et al., *Phys. Rev. Lett.* **126**, 187201 (2021).
34. H. Wang et al., *Phys. Rev. Lett.* **127**, 117202 (2021).
35. T. Devolder et al., *Phys. Rev. B*. **103**, 214431 (2021).
36. A. V. Chumak et al., *Appl. Phys. Lett.* **100**, 082405 (2012).
37. O. d'Allivy Kelly et al., *Appl. Phys. Lett.* **103**, 082408 (2013).
38. A. H. Morrish, *Canted Antiferromagnetism: Hematite* (WORLD SCIENTIFIC, 1995; http://www.worldscientific.com/worldscibooks/10.1142/2518).
39. R. Lebrun et al., *Commun. Phys.* **2**, 50 (2019).
40. V. Vlaminck, M. Bailleul, *Phys. Rev. B*. **81**, 014425 (2010).
41. The Supplementary Material explains the technicalities of the spin-wave spectroscopy measurements using a vector network analyser (S1), including time-gated spin-wave spectroscopy (S2). We also present frequency versus field maps of spin-wave spectroscopy at different k values (S3), and more details measurements on reciprocity and non-reciprocity for k // n and k ⊥ n (S4). We present the theoretical framework to model spin-wave dispersion of bulk and surface spin-wave in colinear and canted antiferromagnets (S5). We then discuss the angular dependency of the absorbed RF power generated by the transducers in the antiferromagnet (S6). We also present magnetometry measurements on the single crystals of hematite (S7), and spin-wave spectroscopy measurements for a magnetic field applied perpendicular to the sample plane (S8).
42. H. J. Fink, *Phys. Rev.* **133**, 1322–1326 (1964).
43. D. E. Beeman, *J. Appl. Phys.* **37**, 1136–1137 (1966).
44. J. Han et al., *Nat. Nanotechnol.* **15**, 563–568 (2020).
45. U. K. Bhaskar, G. Talmelli, F. Ciubotaru, C. Adelmann, T. Devolder, *J. Appl. Phys.* **127**, 033902 (2020).
46. J. Cramer et al., *J. Phys. Appl. Phys.* **51**, 144004 (2018).
47. I. V. Rojdestvenski, M. G. Cottam, A. N. Slavin, *Phys. Rev. B*. **48**, 12768–12777 (1993).
48. M. Jamali, J. H. Kwon, S.-M. Seo, K.-J. Lee, H. Yang, *Sci. Rep.* **3**, 3160 (2013).
49. A. V. Chumak et al., *IEEE Trans. Magn.* **58**, 1–72 (2022).
50. J. Fischer et al., *Phys. Rev. Appl.* **13**, 014019 (2020).
51. M. Althammer et al., *Phys. Rev. B*. **87** (2013), doi:10.1103/PhysRevB.87.224401.
52. H. Wang et al., Long-distance propagation of high-velocity antiferromagnetic spin waves (2022), , doi:10.48550/arXiv.2211.10989.
53. M. Hamdi, F. Posva, D. Grundler, Spin wave dispersion of ultra-low damping hematite ($\alpha\text{-Fe}_2\text{O}_3$) at GHz frequencies (2022), , doi:10.48550/arXiv.2212.11887.

**Acknowledgments**
**Funding:** Financial supports from the Horizon 2020 Framework Programme of the European Commission under FET-Open grant agreement No. 863155 (s-Nebula), under FET-Open grant agreement No. 964931 (TSAR) and under the ITN Grant agreement ID: 861300 (COMRAD) are acknowledged. The authors also acknowledge support from the ANR TRAPIST (ANR-21-CE24-0011).